\documentclass[reprint, amsmath,amssymb,aps,pra, superscriptaddress, hidelinks, floatfix
]{revtex4-2}

\usepackage{graphicx,amsmath}
\usepackage{siunitx}
\usepackage{xstring,hyperref}

\renewcommand{\figurename}{\textbf{Fig.}}

\makeatletter
\renewcommand*{\fnum@figure}{{\normalfont\bfseries \figurename~\thefigure}}
\makeatother

\begin{document}

\title{Observation of slow light in glide-symmetric photonic-crystal waveguides}

\newcommand{\nbi}{Niels Bohr Institute, University of Copenhagen, Blegdamsvej 17, DK-2100 Copenhagen, Denmark}
\newcommand{\fotonik}{Department of Photonics Engineering, DTU Fotonik, Technical University of Denmark, Building 343, DK-2800 Kgs. Lyngby, Denmark}
\newcommand{\icn}{Catalan Institute of Nanoscience and Nanotechnology (ICN2), CSIC and BIST, Campus UAB, Bellaterra, SP-08193 Barcelona, Spain}
\newcommand{\elmore}{Elmore Family School of Electrical and Computer Engineering, Department of Physics and Astronomy, Purdue Quantum Science and Engineering Institute, Purdue University, West Lafayette, IN 47907, USA}
\newcommand{\nanophoton}{NanoPhoton - Center for Nanophotonics, Technical University of Denmark, Ørsteds Plads 345A, DK-2800 Kgs.\ Lyngby, Denmark}

\author{Chirag Murendranath Patil} \affiliation{\nbi} \affiliation{\fotonik}
\author{Guillermo Arregui} \affiliation{\fotonik} \affiliation{\icn}
\author{Morten Mechlenborg} \affiliation{\fotonik}
\author{Xiaoyan Zhou} \affiliation{\nbi}
\author{Hadiseh Alaeian} \affiliation{\elmore}
\author{Pedro David García} \affiliation{\icn}
\author{S{\o}ren Stobbe} \email{ssto@fotonik.dtu.dk} \affiliation{\fotonik} \affiliation{\nanophoton}

\date{\today}

\small

\begin{abstract}
We report optical transmission measurements on suspended silicon photonic-crystal waveguides, where one side of the photonic lattice is shifted by half a period along the waveguide axis. The combination of this glide symmetry and slow light leads to a strongly enhanced chiral light-matter interaction but the interplay between slow light and backscattering has not been investigated experimentally in such waveguides. We build photonic-crystal resonators consisting of glide-symmetric waveguides terminated by reflectors and use transmission measurements as well as evanescent coupling to map out the dispersion relation. We find excellent agreement with theory and measure group indices exceeding 90, implying significant potential for applications in slow-light devices and chiral quantum optics. By measuring resonators of different length, we assess the role of backscattering induced by fabrication imperfections and its intimate connection to the group index.
\end{abstract}


\maketitle 

\section{Introduction}
Dispersion engineering of photonic-crystal waveguides has been a central theme in research on slow light with possible applications in telecommunication photonics such as optical delay lines, optical memories, and optical switches~\cite{okawachi2006all, baba2008slow}. A central objective has been to realize ultra-compact photonic structures with large bandwidths~\cite{krauss2007slow}. Another line of research has focused on exploiting slow-light effects to increase the local density of optical states (LDOS) in order to enhance light-matter interaction towards the manipulation and control of light in quantum photonic systems~\cite{lodahl2015interfacing}. However, the close connection between dispersion, the density of states, and backscattering due to disorder constitutes a strong limitation to practical uses of slow light~\cite{mazoyer2009disorder}. This has led to a number of studies of multiple scattering and Anderson localization in photonic-crystal waveguides~\cite{Vollmer2007experimental,sapienza2010cavity,garcia2017two}, which have evidenced the potential of structural disorder to induce high-$Q$ cavity modes for cavity quantum electrodynamics or random lasing. Notwithstanding these intriguing scientific developments, backscattering and Anderson localization remain the main obstacles for photonic devices employing slow light~\cite{mann_soliton_2017}.\\

More recently, nanophotonic waveguides have been employed in studies of chiral quantum optics~\cite{lodahl2017chiral} where the main figure of merit is the circular LDOS. The transverse spatial confinement of light in conventional waveguides results in positions with local circular polarization~\cite{leFeber2015nanophotonic}, but the polarization tends to be linear at the positions and wavelengths where the LDOS is large. To circumvent this issue, the glide symmetry (GS), i.e., a composition of a mirror symmetry and a translation operation~\cite{mock2010space}, has been shown to be very beneficial, in particular the case in which the photonic lattice on one side of the waveguide core is shifted by half a lattice constant along the waveguide axis~\cite{sollner2015deterministic}. The GS generates local circular polarization at field maxima, which enhances the directional $\beta$-factor. This is a main figure of merit for chiral quantum optics, describing the fraction of emission events ending up in one particular direction and mode relative to the emission into all modes. A directional $\beta \gtrsim 90 \%$ was initially demonstrated in GS waveguides~\cite{sollner2015deterministic}, while geometric tuning of that initial design to induce slow light~\cite{mahmoodian2017engineering} recently led to experimental $\beta \gtrsim 95 \%$, enabling the deterministic generation of path-entangled photon pairs~\cite{pedersen_-demand_2021}.\\
 
 \begin{figure*}[t]
\centering
\includegraphics[width=0.8\textwidth]{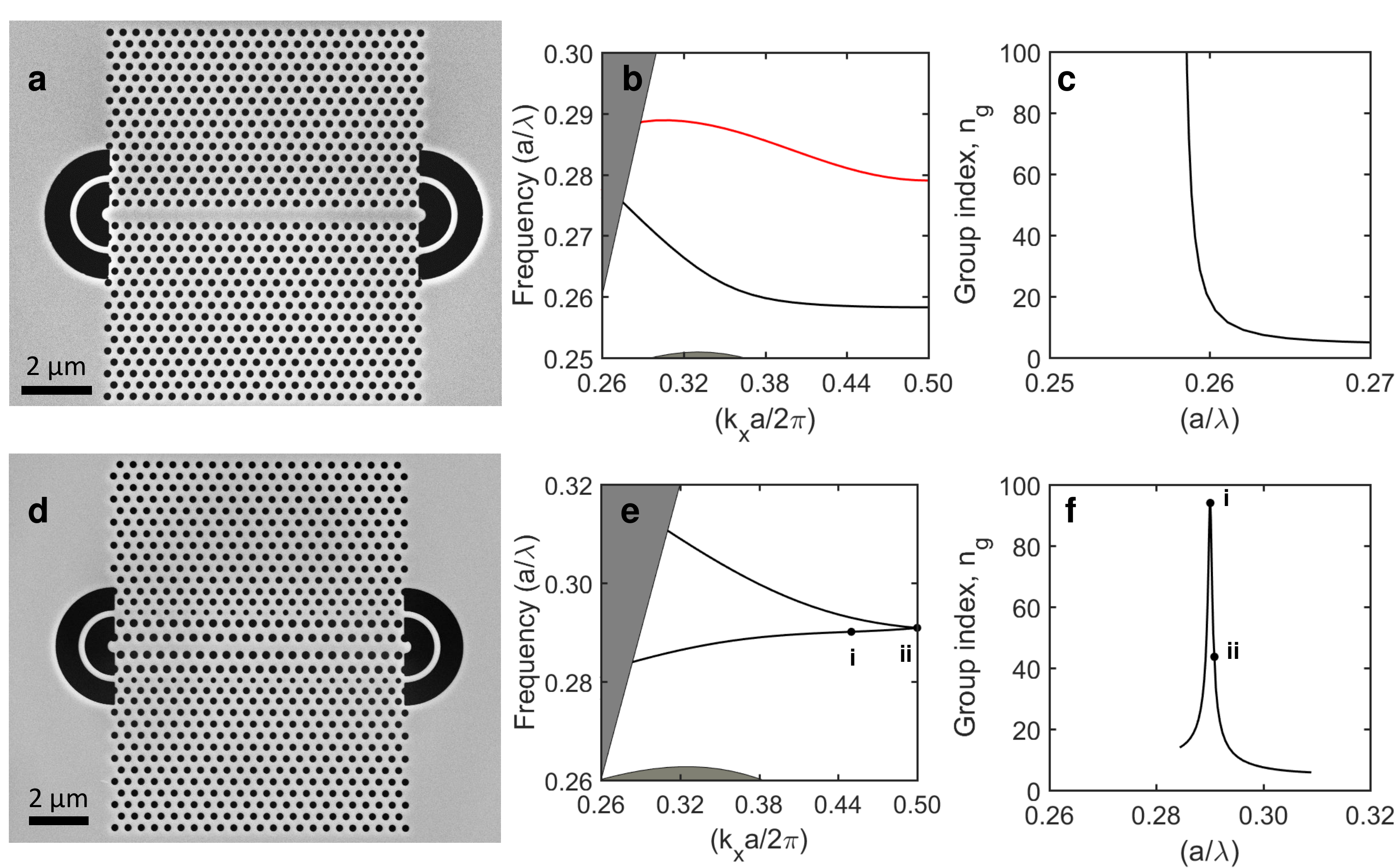}
\caption{Dispersion, slow light, and fabricated resonator devices. (a) Scanning-electron micrograph (SEM) of a \SI{20}{\micro\meter}-long W1 photonic-crystal waveguide terminated by high-reflection grating couplers to form a resonator. (b) Dispersion relation of the W1 waveguide illustrating the transverse-electric-like even (black) and odd (red) waveguide modes. (c) The group index of the even mode is plotted as a function of $a/\lambda$, where $a$ is the lattice constant and $\lambda$ is the wavelength in free space. (d) SEM of a \SI{20}{\micro\meter}-long GS waveguide resonator. (e) Dispersion relation of the GS waveguide exhibiting two modes below the light line. (f) Group index of the GS waveguide. The important points are highlighted as i) the point of maximum group index (${n_g} = 94$) and ii) the zone-edge degeneracy.}
\label{fig:Bandstructure}
\end{figure*}

Further increasing the chiral light-matter coupling calls for operation of GS waveguides at even higher group index, $n_g = c / v_g$, where $c$ is the speed of light in vacuum and $v_g$ is the group velocity. However, an efficient chiral light-matter interface must also minimize the number of backscattered photons due to fabrication imperfection, which typically scales quadratically~\cite{garcia2017two} with $n_g$. This trade-off calls for transmission measurements that experimentally assess the achievable values of $n_g$ and address their connection with disorder-induced backscattering. While previous results on waveguiding in chiral metamaterial waveguides indicated a reduction of backscattering losses in GS waveguides~\cite{orazbayev_chiral_2018,orazbayev_quantitative_2019}, a recent numerical study indicated the opposite for membrane-type GS waveguides~\cite{hauff_chiral_2021}. However, experimental reports on light propagation in GS waveguides are scarce~\cite{sotto2018anomalous,yoshimi_experimental_2021} and have only considered relatively small group indices. In addition, the interplay between the experimentally accessible group index and the sample length has not been investigated. Here we report optical transmission measurements in GS waveguides to map out their dispersion relation, confirming the theoretically predicted dispersion in the slow-light regime, and we analyze the role of fabrication imperfections by comparing the measurements to those in conventional W1 photonic-crystal waveguides. Besides the new insights for chiral quantum optics, our results are important to further studies of GS waveguides, which have been proposed to suppress out-of-plane scattering~\cite{kuang2004reducing} or as a testbed platform to study photonic spin-orbit interaction~\cite{sotto2019spin} and parity-time transitions~\cite{mock2020symmetry}.\\

\section{Design and fabrication of suspended silicon glide-symmetric waveguides}

The GS waveguide we explore follows the design proposed by Mahmoodian et al.~\cite{mahmoodian2017engineering} and exhibits an enhancement of the group index of one order of magnitude relative to previous experiments on GS waveguides~\cite{sollner2015deterministic}. This is achieved by introducing a series of transformations and structural modifications to the geometry of the conventional W1 waveguide shown in Fig.~\ref{fig:Bandstructure}(a). First, the GS is introduced by a half-period shift of one side of the photonic-crystal cladding. The even/odd classification of the modes in the mirror-symmetric W1 waveguide (black and red in Fig.~\ref{fig:Bandstructure}(b)) and the single-mode nature of the even mode with diverging group index~\cite{notomi2001extremely,vlasov2005active} (Fig.~\ref{fig:Bandstructure}(c)) are then lost. In addition, the introduction of the GS directly leads to the formation of pairwise degeneracies of the waveguide bands at the Brillouin zone edge due to the non-symmorphic nature of the GS~\cite{mock2010space}. Second, the three rows of holes closest to the waveguide axis are shifted and their radii are modified, leading to the geometry shown in Fig.~\ref{fig:Bandstructure}(d). This transformation increases the group index at certain points and ensures that the spectral features of interest are within the band gap of the cladding. Importantly, it also ensures that the waveguide is single-moded. The resulting dispersion relation for the transverse-electric-like modes is shown in Fig.~\ref{fig:Bandstructure}(e), which, as Fig.~\ref{fig:Bandstructure}(b), has been obtained with the MPB software package~\cite{MPB}. We note that the maximum group index, $n_\text{g}=94$, (labelled (i) in Figs.~\ref{fig:Bandstructure}(e) and (f)) does not coincide with the degeneracy point at the Brillouin zone edge (labelled (ii) in Figs.~\ref{fig:Bandstructure}(e) and (f)).
We also note that although the degeneracy at the Brillouin zone edge is a salient feature of interface states between topologically different materials~\cite{khanikaev2017two,yoshimi_experimental_2021}, its origin here is rooted in the non-symmorphic character of the GS.\\

\begin{figure*}[t]
\centering\includegraphics[width=0.8\textwidth]{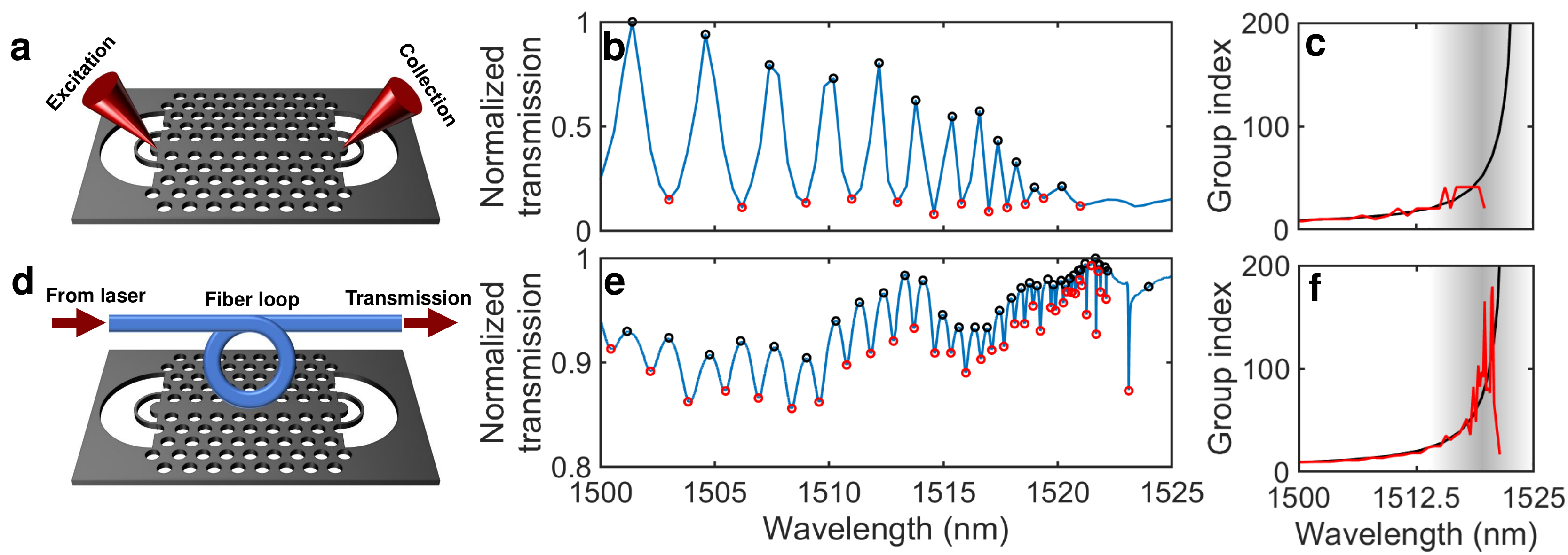}
\caption{Experimental setups and characteristic optical spectra. (a) The resonators are excited by a supercontinuum white-light source, which is directed into and collected from the grating couplers by free-space optics and measured with an optical spectrum analyzer. (b) Transmission spectrum measured in a \SI{70}{\micro\meter}-long W1 photonic-crystal waveguide showing Fabry-Pérot fringe peaks (black circles) and troughs (red circles). (c) Group index extracted from the free spectral range of the resonator device and corrected by a factor of 2 as explained in the main text (red line). The simulated group index is given as a black line. (d) For a second set of measurements we use a tapered fiber loop to evanescently couple to the supported modes using a tunable laser. (e) Transmission spectrum of the same device showing dips at wavelengths where light is evanescently coupled from the fiber loop into the waveguide. (f) Group index extracted from the free spectral range of the measured transmission dips. The correction factor of 2 is not needed in this case. The grey shaded area in (c) and (f) mark the wavelength region where strong backscattering occurs.}
\label{fig:Setup}
\end{figure*}
We have fabricated a series of devices using silicon-on-insulator wafers with a \SI{250}{\nano\meter} silicon device layer and a \SI{3}{\micro\meter} buried oxide layer. The waveguides are directly terminated with circular grating couplers at both ends. These couplers are essentially broadband point scatterers and have a rather low coupling efficiency and high in-plane reflectivity. These properties allow building resonator devices while directing a sizable amount of light out of the chip plane regardless of the waveguide mode profile, which is useful for transmission measurements \cite{faraon2008dipole}. We have carefully measured the clearance dose and applied proximity-effect correction before patterning the devices using \SI{100}{\kilo\electronvolt} electron-beam lithography. The pattern is transferred to the silicon device layer by inductively-coupled plasma reactive-ion etching. Finally, the underlying oxide layer is selectively etched in buffered hydrofluoric acid to realize a membrane structure. We have fabricated in total 60 resonator devices for each waveguide design, consisting of 12 different lengths ($L = \SI{10}{\micro\meter}, \SI{20}{\micro\meter}, ... , \SI{120}{\micro\meter}$) with 5 nominally identical copies of each device. All resonators are fabricated in the same batch, the same chip, and in close proximity to each other, to avoid chip-to-chip variations. This allows statistical analysis of device performance and a direct comparison of the propagation properties of W1 and GS waveguides under the exact same fabrication conditions. Representative scanning electron micrographs of the fabricated devices are shown in Figs.~\ref{fig:Bandstructure}(a) and (d) for W1 and GS waveguides, respectively.\\

\begin{figure*}[t]
\centering
\includegraphics[width=0.7\textwidth]{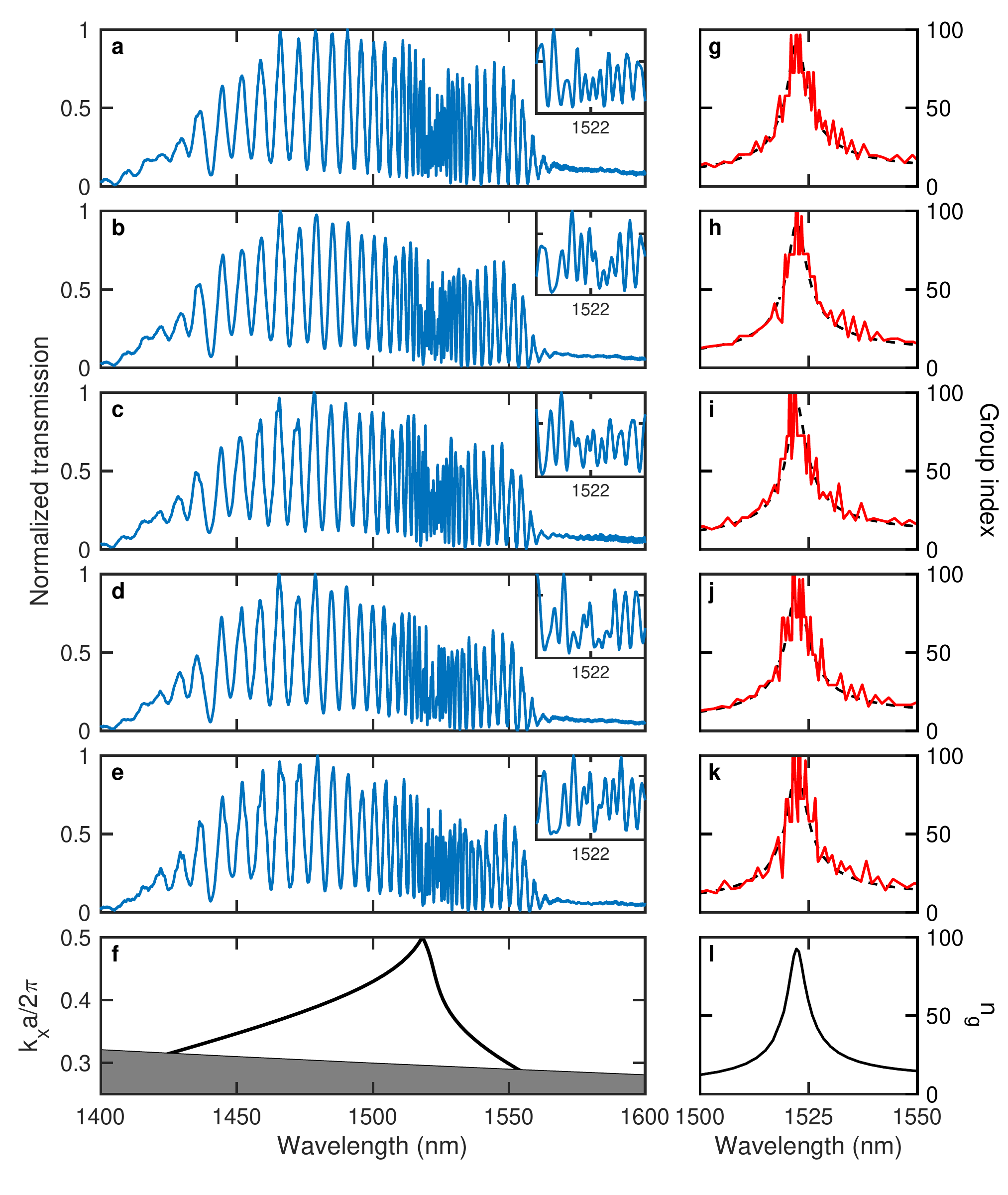}
\caption{Optical transmission measurements for 5 nominally identical GS waveguide resonators of length $L$ = \SI{40}{\micro\meter}. (a-e) The experimental transmission spectra exhibit clear Fabry-Pérot resonances with a monotonously varying free spectral range in agreement with (f) the calculated dispersion relation. (g-k) The extracted group index agrees very well with (l) theory.}
\label{fig:TransmissionSpectra_GPW_Chirag_Copies}
\end{figure*}

\section{Transmission measurements}
We characterize the devices using free-space optical transmission measurements performed with the experimental setup shown in Fig.~\ref{fig:Setup}(a). A supercontinuum coherent white-light source with long-pass filtering is used to obtain a broad input spectrum (\SI{1200}{\nano\meter} to \SI{1700}{\nano\meter}). The excitation light is directed onto one of the grating couplers on each device using a confocal setup with polarization control to ensure that only TE-like modes are excited. The light transmitted through the cavity is extracted from the second grating coupler and characterized using an optical spectrum analyzer with a spectral resolution of \SI{0.5}{\nano\meter}. A characteristic measurement on a W1 waveguide is shown in Fig.~\ref{fig:Setup}(b). From the observed Fabry-Pérot (FP) fringes, the group index is extracted as
\begin{equation}
n_g = \frac{\lambda^2}{2 \Delta\lambda_\text{FSR} L},
\label{eqn:FSR}
\end{equation}
where $\lambda$ is the wavelength, $\Delta\lambda_\text{FSR}$ is the free spectral range between consecutive FP cavity modes and $L$ is the resonator length. The group index obtained from an automatic peak-finding algorithm is shown in red in Fig.~\ref{fig:Setup}(c), showing good agreement with the theoretical curve in solid black. We note, however, that two transformations are applied. Firstly, the calculated dispersion relation and group index are shifted by an offset of \SI{22}{\nano\meter}, which compensates for a systematic uncertainty in the hole radii and slab thickness. Secondly, the experimentally extracted group index is multiplied by a factor of 2. This is needed because of the grating couplers, which may allow coupling into multiple spatial positions and thus resulting in an interference with only every second oscillation being resolved, which is also visible in the raw data in Fig.~\ref{fig:Setup}(b) as an anharmonic component in the FP oscillations. This is further confirmed by a windowed numerical Fourier transform analysis of the raw data as well as experiments using an alternative coupling scheme, as discussed in further detail below. However, since the waveguide is highly dispersive, we resort to an automatic peak search and systematically apply the factor of two afterwards. To double-check that this factor originates from the coupling scheme, a second setup, illustrated in Fig.~\ref{fig:Setup}(d), is used. In this case, a tapered fiber loop is used to evanescently couple light into the waveguides \cite{lee2008characterizing}, where the coupling to the cavity modes results in transmission dips, as shown in Fig.~\ref{fig:Setup}(e). A direct comparison of Figs.~\ref{fig:Setup}(b) and (e) in the wavelength range 1500-1520 nm evidences that the fiber allows coupling to all FP modes without any interference effects and confirming that the destructive interference of every second mode observed in Fig.~\ref{fig:Setup}(b) is due to the coupling condition and not the waveguide. Above 1520 nm, the evanescent technique reveals additional resonances at random wavelengths above the cut-off observed in the free-space measurement. The fluctuations in the free spectral range lead to strong fluctuations in the extracted value of $n_g$ around the theoretical curve (Fig.~\ref{fig:Setup}(f)). Such resonances originate from the enhanced backscattering induced by fabrication imperfection in the slow-light regime which ultimately induces Anderson localization~\cite{Vollmer2007experimental,sapienza2010cavity,garcia2017two}. They correspond to tightly localized modes that are generally optically dark when employing the free-space grating couplers but can be evidenced with the local evanescent probe~\cite{arregui_all-optical_2018}, thus explaining the difference between the two spectra. Despite the high quality of these evanescent measurements and the additional insights they provide regarding coherent multiple scattering in the device under test, the symmetry and phase-matching selection rules~\cite{mock2010space} do not allow coupling to the higher-energy band of this GS waveguide using the fiber taper. We therefore limit the analysis herein to free-space measurements and apply a factor of 2 in the extracted $n_g$ to all measured resonators.

\begin{figure*}[t]
\centering
\includegraphics[width=0.7\textwidth]{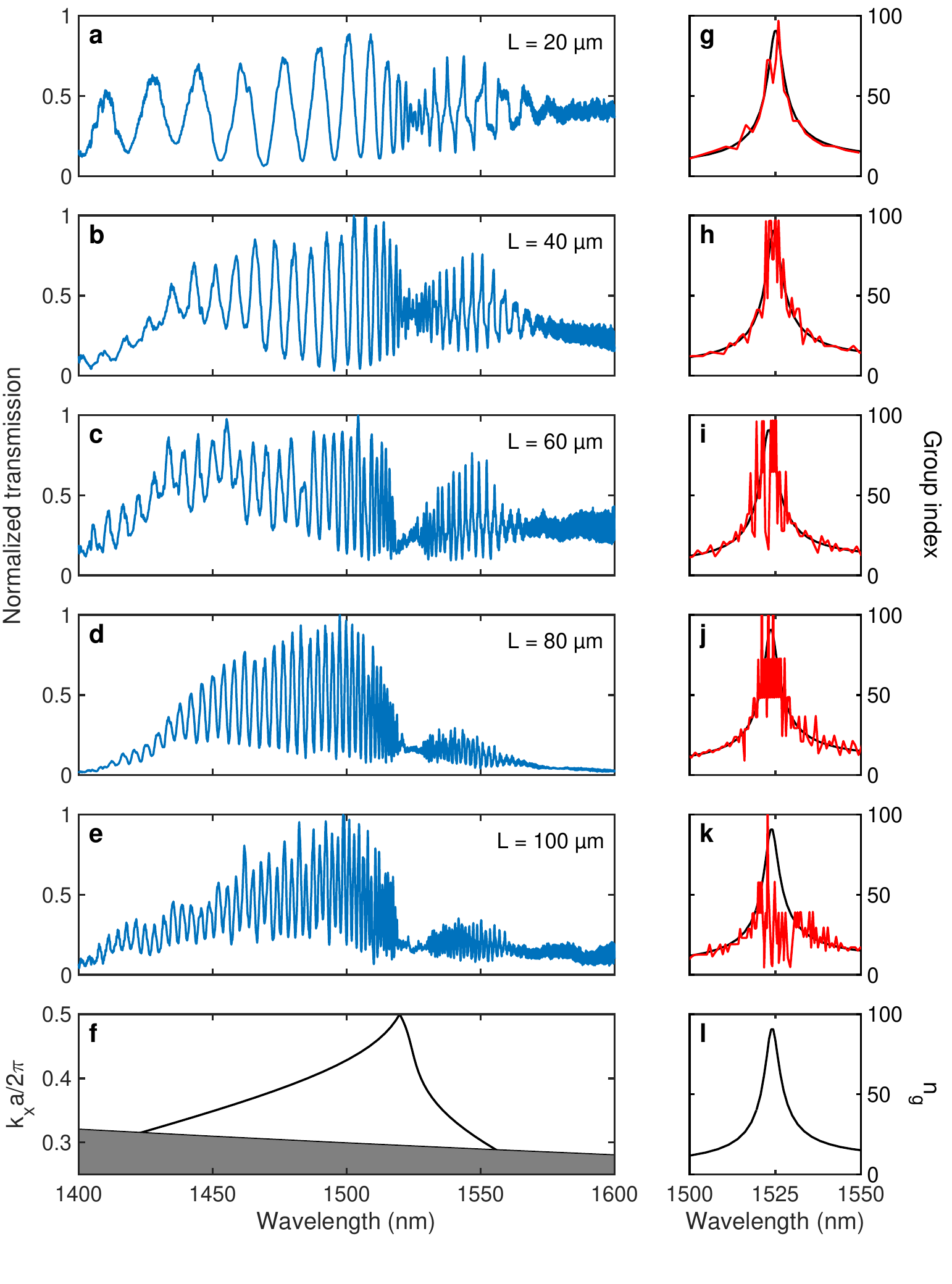}
\caption{Optical transmission measurements for GS waveguide resonators of different length. (a-e) Normalized raw data for resonator lengths varying from \SI{20}{\micro\meter} to \SI{100}{\micro\meter}, which show a pronounced reduction of the transmission around the group-index maximum in (f) the dispersion relation. The limited signal-to-noise ratio and Anderson localization effects leads to substantial variations in the extracted group index for longer devices but the agreement between (g-k) experiment and (l) theory is excellent for $L\leq\SI{40}{\micro\meter}$.}
\label{fig:TransmissionSpectra_GPW_Chirag_L}
\end{figure*}

\begin{figure*}[t]
\centering
\includegraphics[width=0.7\textwidth]{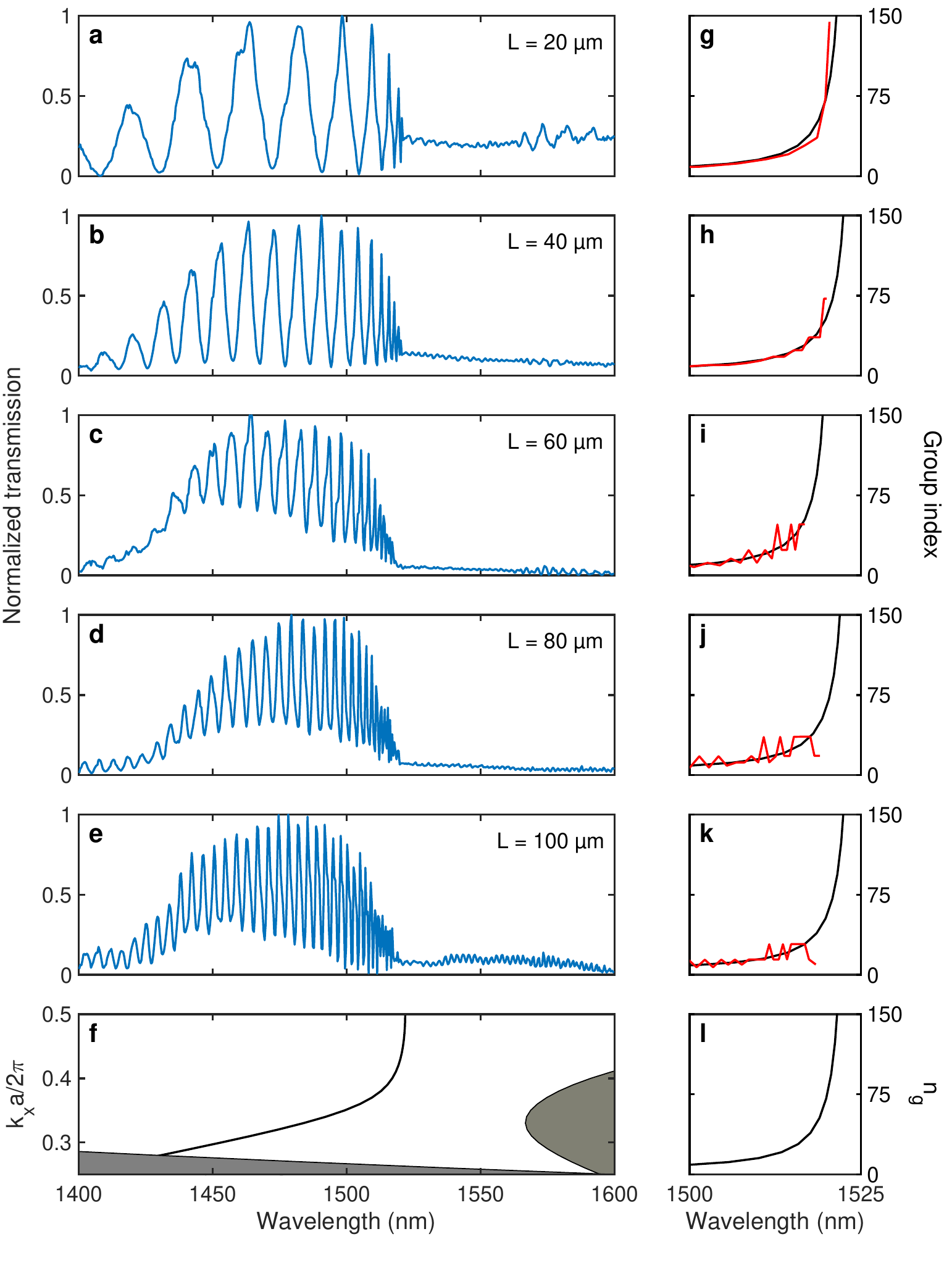}
\caption{Optical transmission measurements for W1 waveguide resonators of different length. (a-e) The experimental transmission spectra for different lengths exhibit clear Fabry-Pérot resonances and a cut-off in agreement with (f) the calculated dispersion relation. (g-k) The extracted group index agrees very well with (l) theory below the threshold for Anderson localization.}
\label{fig:TransmissionSpectra_W1_Chirag_L}
\end{figure*}

Figure \ref{fig:TransmissionSpectra_GPW_Chirag_Copies} shows normalized raw data for five nominally identical and short ($L$ = \SI{40}{\micro\meter}) GS waveguides along with the extracted group indices. In this case, the calculated dispersion and group index, shown in Figs.~\ref{fig:TransmissionSpectra_GPW_Chirag_Copies}(f) and (l), have been shifted by 23 nm. The agreement between theory and experiment is generally excellent and the fluctuations in the measured group indices arise primarily from the limited analyzer resolution and the signal-to-noise ratio, in particular around the slow-light maximum (see insets in Figs.~\ref{fig:TransmissionSpectra_GPW_Chirag_Copies}(a)-(e)). In addition, minor fluctuations in the group index can also be attributed to dispersion in the optical feedback of the resonator~\cite{sarkissian2014group}, here provided by the grating couplers. As a major result of our work, we experimentally confirm the theoretically predicted group indices exceeding $n_\text{g}=90$, a value twice as large as that previously measured in a GS waveguide~\cite{yoshimi_experimental_2021}.\\

\begin{figure*}[t]
\centering
\includegraphics[width=0.6\textwidth]{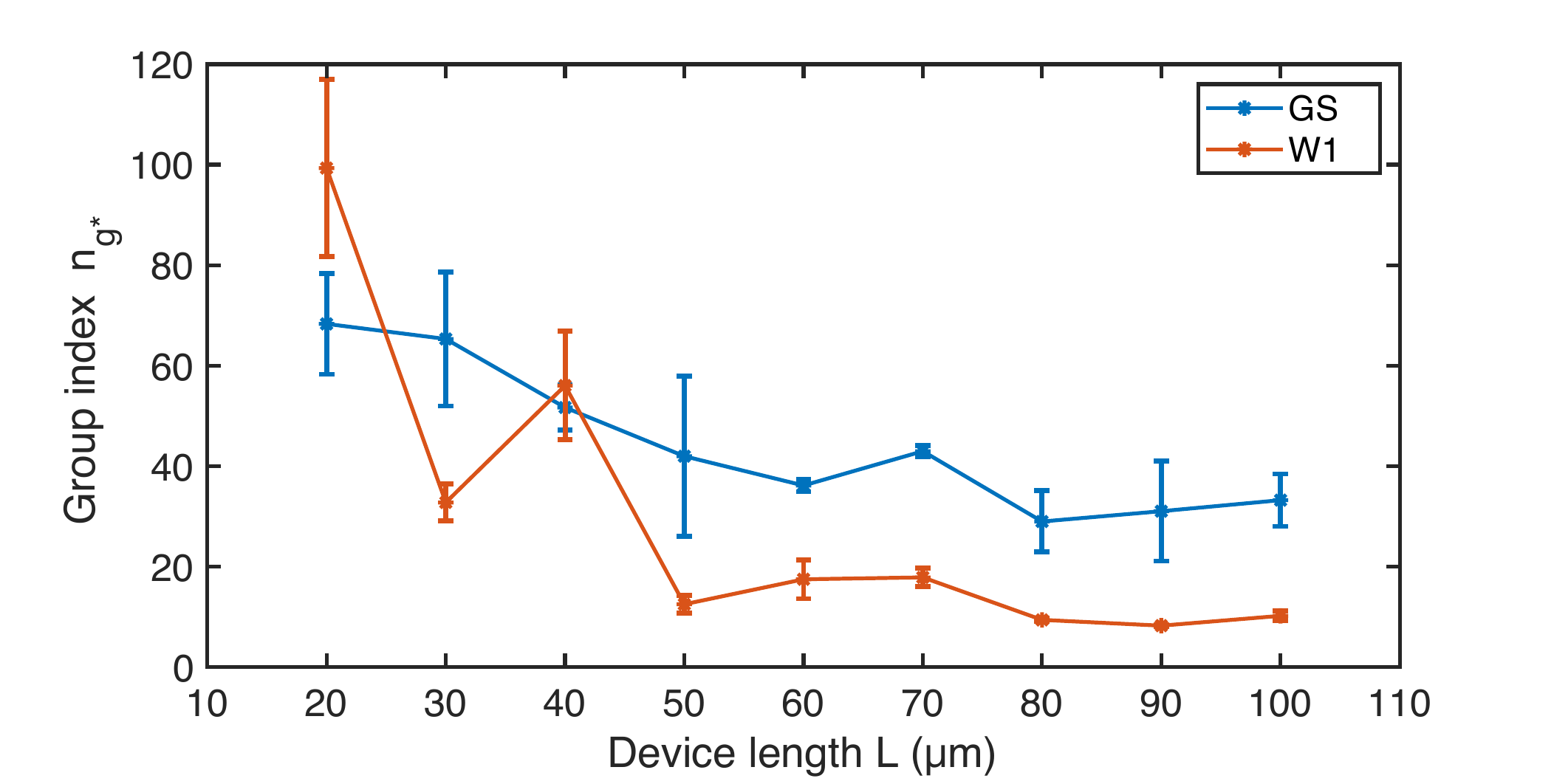}
\caption{Measured ensemble-averaged group index at the cut-off wavelength as a function of device length for W1 (red) and GS waveguides (blue). The error bars correspond to the statistical standard deviation obtained from 5 nominally identical resonators. The GS waveguide systematically sustains higher group indices for device lengths $L > \SI{50}{\micro\meter}$.}
\label{fig:ng_cutoff}
\end{figure*}

To assess the feasibility of chiral quantum optics experiments in the slow-light regime of the GS waveguide described here, we carry out transmission measurements for different waveguide lengths (Fig.~\ref{fig:TransmissionSpectra_GPW_Chirag_L}). For the shorter devices, as in Fig.~\ref{fig:TransmissionSpectra_GPW_Chirag_Copies}, we observe an excellent agreement between the extracted group index and the theoretical predictions. For longer devices, the use of Eq.~(\ref{eqn:FSR}), which relies on the assumption that light can travel back and forth freely, fails to reproduce the calculated curve above a certain value of $n_g$. The enhanced losses in this region lead to an increasingly limited signal-to-noise ratio, and backscattering ultimately induces Anderson localization of the light field at random wavelengths. The free spectral range between the observed peaks is thus affected and Eq.~(\ref{eqn:FSR}) deviates considerably from the numerically calculated curve. We also carry out reference experiments on a set of conventional W1 waveguides fabricated on the same chip to ensure that the structural disorder is statistically identical. Characteristic measurements are shown in Fig.~\ref{fig:TransmissionSpectra_W1_Chirag_L}, with FP resonances observed for all sample lengths and with an envelope transmission dropping rapidly near the band edge. The experimentally extracted group indices show good agreement with theory up to a clear cut-off wavelength, which blueshifts with increasing device length. Unlike for the W1 waveguides, the GS waveguides do not exhibit this sharp cut-off on both sides of the $n_g$ maximum but multiple low-amplitude transmission peaks are detected throughout the full wavelength range in all measured devices.\\

We extend the measurements shown in Figs.~\ref{fig:TransmissionSpectra_GPW_Chirag_L} and \ref{fig:TransmissionSpectra_W1_Chirag_L} to all nominally identical devices in order to assess both types of waveguides statistically and to extract a critical length above which they cannot be safely operated in the slow-light regime with a given slowdown factor $n_g$. Since W1 and GS waveguides have nearly identical features (circular holes with similar radii) and are fabricated in parallel on the same chip, their structural disorder can be assumed uniform, which allows their performance to be compared at a common and realistic disorder level and group index~\cite{arregui_quantifying_2021}. We define, for a fixed length $L$, a cut-off operational point, $\{\lambda_*,n_{\text{g}*}\}$, at the wavelength $\lambda_*$ at which the reconstructed $n_{\text{g}}$ deviates a specified amount from the theoretical curve, an approach that can be systematically applied to both W1 and GS waveguides. Specifically, we consider the normalized error,
\begin{equation}
\delta(\lambda) = \frac{|n_\text{g}^\text{exp}(\lambda) - n_\text{g}^\text{the}(\lambda)|}{n_\text{g}^\text{exp}(\lambda) + n_\text{g}^\text{the}(\lambda)},
\label{eqn:cutoff_criterion}
\end{equation}
where "$\text{the}$" and "$\text{exp}$" denote theory (without disorder) and experiment, respectively. The cut-off wavelength $\lambda_*$ is defined by $\delta(\lambda_*)>\delta_*$ or the wavelength corresponding to the maximum $n_g$ measured in case the criteria is not satisfied within the measured data. The precise value of the threshold $\delta_*$ influences the extracted pair $\{\lambda_*,n_{\text{g}*}\}$ but setting this value to, e.g. $\delta_*=0.2$, and assessing the deviations statistically (using the 5 nominally-identical devices) allows a fair comparative analysis of W1 and GS waveguides for a fixed device length. Figure \ref{fig:ng_cutoff} shows the average $n_{g*}$ as a function of the device length $L$ for both types of waveguides, with the error bars representing the standard deviation among the 5 resonators. The negative correlation between group index and losses typical of conventional W1 waveguides~\cite{garcia2010density} is also found for GS waveguides. However, the higher maximum group indices $n_{\text{g}*}$ of GS waveguides relative to W1 waveguides for lengths $L$ = 50-100 \SI{}{\micro\meter} indicate lower propagation losses at high group indices. This experimental observation may seem in disagreement with recent perturbation theory calculations~\cite{hauff_chiral_2021}, but the r.m.s.\ roughness level we observe in our devices is smaller than the 3 nm considered therein, even in absolute terms but even smaller relative to the wavelength, and their calculations do not account for coherent multiple scattering effects and radiation losses.

\section{Conclusion}
Our measurements confirm theoretical predictions of the dispersion relation and group index of waveguides with GS. Importantly, we measured group indices exceeding $n_\text{g}=90$, which has important implications for research on chiral quantum optics because our results imply experimentally achievable directional $\beta$-factors approaching unity, even beyond what has been recently reported in Ref.~\cite{pedersen_-demand_2021}. We have addressed the dispersive propagation losses of a GS waveguide by evaluating fluctuations in the extracted group index as a function of the resonator length. Comparison of these deviations to those observed in conventional W1 waveguides over a set of nominally identical devices indicate that GS waveguides offer a competitive advantage when operated at group indices below 50, effectively providing delay times twice as large as those achieved in W1 waveguides. Beyond the differences observed between the two waveguide designs, the methodology used here constitutes an important step forward in understanding light transport in photonic-crystal waveguides subject to structural disorder, which in the slow-light regime is governed by the localization length, $\xi$. While the exponential decay in transmission measurements cannot disentangle backscattering losses from other losses~\cite{sapienza2010cavity}, such as absorption or out-of-plane scattering, the spectral properties of the (quasi-)normal modes in an open finite system~\cite{mondal_optical_2019} are better suited to obtain $\xi$. The Thouless criterion~\cite{edwards1972numerical} is not directly applicable to the cavities here, which exhibit transmission peaks even in the absence of scattering. A more sophisticated statistical study of the group-index fluctuations we observe may be used to extract $\xi$, similar to what has been done for two-dimensional systems~\cite{mondal_relation_2020}, where a univoque relation between the localization length and the level-spacing probability distribution has been predicted.

\section*{Funding}
We gratefully acknowledge financial support from the Villum Foundation Young Investigator Program (Grant No.\ 13170), the Danish National Research Foundation (Grant No.\ DNRF147 - NanoPhoton), Innovation Fund Denmark (Grant No.\ 0175-00022 - NEXUS), and the Independent Research Fund Denmark (Grant No.\ 0135-00315 - VAFL).



\bibliographystyle{apsrev4-2}
\bibliography{main}

\begin{thebibliography}{36}%
\makeatletter
\providecommand \@ifxundefined [1]{%
 \@ifx{#1\undefined}
}%
\providecommand \@ifnum [1]{%
 \ifnum #1\expandafter \@firstoftwo
 \else \expandafter \@secondoftwo
 \fi
}%
\providecommand \@ifx [1]{%
 \ifx #1\expandafter \@firstoftwo
 \else \expandafter \@secondoftwo
 \fi
}%
\providecommand \natexlab [1]{#1}%
\providecommand \enquote  [1]{``#1''}%
\providecommand \bibnamefont  [1]{#1}%
\providecommand \bibfnamefont [1]{#1}%
\providecommand \citenamefont [1]{#1}%
\providecommand \href@noop [0]{\@secondoftwo}%
\providecommand \href [0]{\begingroup \@sanitize@url \@href}%
\providecommand \@href[1]{\@@startlink{#1}\@@href}%
\providecommand \@@href[1]{\endgroup#1\@@endlink}%
\providecommand \@sanitize@url [0]{\catcode `\\12\catcode `\$12\catcode
  `\&12\catcode `\#12\catcode `\^12\catcode `\_12\catcode `\%12\relax}%
\providecommand \@@startlink[1]{}%
\providecommand \@@endlink[0]{}%
\providecommand \url  [0]{\begingroup\@sanitize@url \@url }%
\providecommand \@url [1]{\endgroup\@href {#1}{\urlprefix }}%
\providecommand \urlprefix  [0]{URL }%
\providecommand \Eprint [0]{\href }%
\providecommand \doibase [0]{https://doi.org/}%
\providecommand \selectlanguage [0]{\@gobble}%
\providecommand \bibinfo  [0]{\@secondoftwo}%
\providecommand \bibfield  [0]{\@secondoftwo}%
\providecommand \translation [1]{[#1]}%
\providecommand \BibitemOpen [0]{}%
\providecommand \bibitemStop [0]{}%
\providecommand \bibitemNoStop [0]{.\EOS\space}%
\providecommand \EOS [0]{\spacefactor3000\relax}%
\providecommand \BibitemShut  [1]{\csname bibitem#1\endcsname}%
\let\auto@bib@innerbib\@empty
\bibitem [{\citenamefont {Okawachi}\ \emph {et~al.}(2006)\citenamefont
  {Okawachi}, \citenamefont {Foster}, \citenamefont {Sharping}, \citenamefont
  {Gaeta}, \citenamefont {Xu},\ and\ \citenamefont {Lipson}}]{okawachi2006all}%
  \BibitemOpen
  \bibfield  {author} {\bibinfo {author} {\bibfnamefont {Y.}~\bibnamefont
  {Okawachi}}, \bibinfo {author} {\bibfnamefont {M.~A.}\ \bibnamefont
  {Foster}}, \bibinfo {author} {\bibfnamefont {J.~E.}\ \bibnamefont
  {Sharping}}, \bibinfo {author} {\bibfnamefont {A.~L.}\ \bibnamefont {Gaeta}},
  \bibinfo {author} {\bibfnamefont {Q.}~\bibnamefont {Xu}},\ and\ \bibinfo
  {author} {\bibfnamefont {M.}~\bibnamefont {Lipson}},\ }\href
  {https://doi.org/10.1364/OE.14.002317} {\bibfield  {journal} {\bibinfo
  {journal} {Opt. Express}\ }\textbf {\bibinfo {volume} {14}},\ \bibinfo
  {pages} {2317} (\bibinfo {year} {2006})}\BibitemShut {NoStop}%
\bibitem [{\citenamefont {Baba}(2008)}]{baba2008slow}%
  \BibitemOpen
  \bibfield  {author} {\bibinfo {author} {\bibfnamefont {T.}~\bibnamefont
  {Baba}},\ }\href {https://doi.org/10.1038/nphoton.2008.146} {\bibfield
  {journal} {\bibinfo  {journal} {Nat. Photon.}\ }\textbf {\bibinfo {volume}
  {2}},\ \bibinfo {pages} {465} (\bibinfo {year} {2008})}\BibitemShut {NoStop}%
\bibitem [{\citenamefont {Krauss}(2007)}]{krauss2007slow}%
  \BibitemOpen
  \bibfield  {author} {\bibinfo {author} {\bibfnamefont {T.~F.}\ \bibnamefont
  {Krauss}},\ }\href {https://doi.org/10.1088/0022-3727/40/9/S07} {\bibfield
  {journal} {\bibinfo  {journal} {J. Phys. D}\ }\textbf {\bibinfo {volume}
  {40}},\ \bibinfo {pages} {2666} (\bibinfo {year} {2007})}\BibitemShut
  {NoStop}%
\bibitem [{\citenamefont {Lodahl}\ \emph {et~al.}(2015)\citenamefont {Lodahl},
  \citenamefont {Mahmoodian},\ and\ \citenamefont
  {Stobbe}}]{lodahl2015interfacing}%
  \BibitemOpen
  \bibfield  {author} {\bibinfo {author} {\bibfnamefont {P.}~\bibnamefont
  {Lodahl}}, \bibinfo {author} {\bibfnamefont {S.}~\bibnamefont {Mahmoodian}},\
  and\ \bibinfo {author} {\bibfnamefont {S.}~\bibnamefont {Stobbe}},\ }\href
  {https://doi.org/10.1103/RevModPhys.87.347} {\bibfield  {journal} {\bibinfo
  {journal} {Reviews of Modern Physics}\ }\textbf {\bibinfo {volume} {87}},\
  \bibinfo {pages} {347} (\bibinfo {year} {2015})}\BibitemShut {NoStop}%
\bibitem [{\citenamefont {Mazoyer}\ \emph {et~al.}(2009)\citenamefont
  {Mazoyer}, \citenamefont {Hugonin},\ and\ \citenamefont
  {Lalanne}}]{mazoyer2009disorder}%
  \BibitemOpen
  \bibfield  {author} {\bibinfo {author} {\bibfnamefont {S.}~\bibnamefont
  {Mazoyer}}, \bibinfo {author} {\bibfnamefont {J.~P.}\ \bibnamefont
  {Hugonin}},\ and\ \bibinfo {author} {\bibfnamefont {P.}~\bibnamefont
  {Lalanne}},\ }\href@noop {} {\bibfield  {journal} {\bibinfo  {journal} {Phys.
  Rev. Lett}\ }\textbf {\bibinfo {volume} {103}},\ \bibinfo {pages} {063903}
  (\bibinfo {year} {2009})}\BibitemShut {NoStop}%
\bibitem [{\citenamefont {Topolancik}\ \emph {et~al.}(2007)\citenamefont
  {Topolancik}, \citenamefont {Ilic},\ and\ \citenamefont
  {Vollmer}}]{Vollmer2007experimental}%
  \BibitemOpen
  \bibfield  {author} {\bibinfo {author} {\bibfnamefont {J.}~\bibnamefont
  {Topolancik}}, \bibinfo {author} {\bibfnamefont {B.}~\bibnamefont {Ilic}},\
  and\ \bibinfo {author} {\bibfnamefont {F.}~\bibnamefont {Vollmer}},\ }\href
  {https://doi.org/10.1103/PhysRevLett.99.253901} {\bibfield  {journal}
  {\bibinfo  {journal} {Phys. Rev. Lett.}\ }\textbf {\bibinfo {volume} {99}},\
  \bibinfo {pages} {253901} (\bibinfo {year} {2007})}\BibitemShut {NoStop}%
\bibitem [{\citenamefont {Sapienza}\ \emph {et~al.}(2010)\citenamefont
  {Sapienza}, \citenamefont {Thyrrestrup}, \citenamefont {Stobbe},
  \citenamefont {Garcia}, \citenamefont {Smolka},\ and\ \citenamefont
  {Lodahl}}]{sapienza2010cavity}%
  \BibitemOpen
  \bibfield  {author} {\bibinfo {author} {\bibfnamefont {L.}~\bibnamefont
  {Sapienza}}, \bibinfo {author} {\bibfnamefont {H.}~\bibnamefont
  {Thyrrestrup}}, \bibinfo {author} {\bibfnamefont {S.}~\bibnamefont {Stobbe}},
  \bibinfo {author} {\bibfnamefont {P.~D.}\ \bibnamefont {Garcia}}, \bibinfo
  {author} {\bibfnamefont {S.}~\bibnamefont {Smolka}},\ and\ \bibinfo {author}
  {\bibfnamefont {P.}~\bibnamefont {Lodahl}},\ }\href
  {https://doi.org/10.1126/science.1185080} {\bibfield  {journal} {\bibinfo
  {journal} {Science}\ }\textbf {\bibinfo {volume} {327}},\ \bibinfo {pages}
  {1352} (\bibinfo {year} {2010})}\BibitemShut {NoStop}%
\bibitem [{\citenamefont {Garc{\'\i}a}\ \emph {et~al.}(2017)\citenamefont
  {Garc{\'\i}a}, \citenamefont {Kir{\v{s}}ansk{\.e}}, \citenamefont {Javadi},
  \citenamefont {Stobbe},\ and\ \citenamefont {Lodahl}}]{garcia2017two}%
  \BibitemOpen
  \bibfield  {author} {\bibinfo {author} {\bibfnamefont {P.~D.}\ \bibnamefont
  {Garc{\'\i}a}}, \bibinfo {author} {\bibfnamefont {G.}~\bibnamefont
  {Kir{\v{s}}ansk{\.e}}}, \bibinfo {author} {\bibfnamefont {A.}~\bibnamefont
  {Javadi}}, \bibinfo {author} {\bibfnamefont {S.}~\bibnamefont {Stobbe}},\
  and\ \bibinfo {author} {\bibfnamefont {P.}~\bibnamefont {Lodahl}},\ }\href
  {https://doi.org/10.1103/PhysRevB.96.144201} {\bibfield  {journal} {\bibinfo
  {journal} {Phys. Rev. B}\ }\textbf {\bibinfo {volume} {96}},\ \bibinfo
  {pages} {144201} (\bibinfo {year} {2017})}\BibitemShut {NoStop}%
\bibitem [{\citenamefont {Mann}\ and\ \citenamefont
  {Hughes}(2017)}]{mann_soliton_2017}%
  \BibitemOpen
  \bibfield  {author} {\bibinfo {author} {\bibfnamefont {N.}~\bibnamefont
  {Mann}}\ and\ \bibinfo {author} {\bibfnamefont {S.}~\bibnamefont {Hughes}},\
  }\href {https://doi.org/10.1103/PhysRevLett.118.253901} {\bibfield  {journal}
  {\bibinfo  {journal} {Physical Review Letters}\ }\textbf {\bibinfo {volume}
  {118}},\ \bibinfo {pages} {253901} (\bibinfo {year} {2017})}\BibitemShut
  {NoStop}%
\bibitem [{\citenamefont {Lodahl}\ \emph {et~al.}(2017)\citenamefont {Lodahl},
  \citenamefont {Mahmoodian}, \citenamefont {Stobbe}, \citenamefont
  {Rauschenbeutel}, \citenamefont {Schneeweiss}, \citenamefont {Volz},
  \citenamefont {Pichler},\ and\ \citenamefont {Zoller}}]{lodahl2017chiral}%
  \BibitemOpen
  \bibfield  {author} {\bibinfo {author} {\bibfnamefont {P.}~\bibnamefont
  {Lodahl}}, \bibinfo {author} {\bibfnamefont {S.}~\bibnamefont {Mahmoodian}},
  \bibinfo {author} {\bibfnamefont {S.}~\bibnamefont {Stobbe}}, \bibinfo
  {author} {\bibfnamefont {A.}~\bibnamefont {Rauschenbeutel}}, \bibinfo
  {author} {\bibfnamefont {P.}~\bibnamefont {Schneeweiss}}, \bibinfo {author}
  {\bibfnamefont {J.}~\bibnamefont {Volz}}, \bibinfo {author} {\bibfnamefont
  {H.}~\bibnamefont {Pichler}},\ and\ \bibinfo {author} {\bibfnamefont
  {P.}~\bibnamefont {Zoller}},\ }\href {https://doi.org/10.1038/nature21037}
  {\bibfield  {journal} {\bibinfo  {journal} {Nature}\ }\textbf {\bibinfo
  {volume} {541}},\ \bibinfo {pages} {473} (\bibinfo {year}
  {2017})}\BibitemShut {NoStop}%
\bibitem [{\citenamefont {Le~Feber}\ \emph {et~al.}(2015)\citenamefont
  {Le~Feber}, \citenamefont {Rotenberg},\ and\ \citenamefont
  {Kuipers}}]{leFeber2015nanophotonic}%
  \BibitemOpen
  \bibfield  {author} {\bibinfo {author} {\bibfnamefont {B.}~\bibnamefont
  {Le~Feber}}, \bibinfo {author} {\bibfnamefont {N.}~\bibnamefont
  {Rotenberg}},\ and\ \bibinfo {author} {\bibfnamefont {L.}~\bibnamefont
  {Kuipers}},\ }\href {https://doi.org/10.1038/ncomms7695} {\bibfield
  {journal} {\bibinfo  {journal} {Nat. Commun.}\ }\textbf {\bibinfo {volume}
  {6}},\ \bibinfo {pages} {6695} (\bibinfo {year} {2015})}\BibitemShut
  {NoStop}%
\bibitem [{\citenamefont {Mock}\ \emph {et~al.}(2010)\citenamefont {Mock},
  \citenamefont {Lu},\ and\ \citenamefont {O’Brien}}]{mock2010space}%
  \BibitemOpen
  \bibfield  {author} {\bibinfo {author} {\bibfnamefont {A.}~\bibnamefont
  {Mock}}, \bibinfo {author} {\bibfnamefont {L.}~\bibnamefont {Lu}},\ and\
  \bibinfo {author} {\bibfnamefont {J.}~\bibnamefont {O’Brien}},\ }\href
  {https://doi.org/10.1103/PhysRevB.81.155115} {\bibfield  {journal} {\bibinfo
  {journal} {Phys. Rev. B}\ }\textbf {\bibinfo {volume} {81}},\ \bibinfo
  {pages} {155115} (\bibinfo {year} {2010})}\BibitemShut {NoStop}%
\bibitem [{\citenamefont {S{\"o}llner}\ \emph {et~al.}(2015)\citenamefont
  {S{\"o}llner}, \citenamefont {Mahmoodian}, \citenamefont {Hansen},
  \citenamefont {Midolo}, \citenamefont {Javadi}, \citenamefont
  {Kir{\v{s}}ansk{\.e}}, \citenamefont {Pregnolato}, \citenamefont {El-Ella},
  \citenamefont {Lee}, \citenamefont {Song}, \citenamefont {Stobbe},\ and\
  \citenamefont {Lodahl}}]{sollner2015deterministic}%
  \BibitemOpen
  \bibfield  {author} {\bibinfo {author} {\bibfnamefont {I.}~\bibnamefont
  {S{\"o}llner}}, \bibinfo {author} {\bibfnamefont {S.}~\bibnamefont
  {Mahmoodian}}, \bibinfo {author} {\bibfnamefont {S.~L.}\ \bibnamefont
  {Hansen}}, \bibinfo {author} {\bibfnamefont {L.}~\bibnamefont {Midolo}},
  \bibinfo {author} {\bibfnamefont {A.}~\bibnamefont {Javadi}}, \bibinfo
  {author} {\bibfnamefont {G.}~\bibnamefont {Kir{\v{s}}ansk{\.e}}}, \bibinfo
  {author} {\bibfnamefont {T.}~\bibnamefont {Pregnolato}}, \bibinfo {author}
  {\bibfnamefont {H.}~\bibnamefont {El-Ella}}, \bibinfo {author} {\bibfnamefont
  {E.~H.}\ \bibnamefont {Lee}}, \bibinfo {author} {\bibfnamefont {J.~D.}\
  \bibnamefont {Song}}, \bibinfo {author} {\bibfnamefont {S.}~\bibnamefont
  {Stobbe}},\ and\ \bibinfo {author} {\bibfnamefont {P.}~\bibnamefont
  {Lodahl}},\ }\href {https://doi.org/10.1038/nnano.2015.159} {\bibfield
  {journal} {\bibinfo  {journal} {Nat. Nanotechnol.}\ }\textbf {\bibinfo
  {volume} {10}},\ \bibinfo {pages} {775} (\bibinfo {year} {2015})}\BibitemShut
  {NoStop}%
\bibitem [{\citenamefont {Mahmoodian}\ \emph {et~al.}(2017)\citenamefont
  {Mahmoodian}, \citenamefont {Prindal-Nielsen}, \citenamefont {S{\"o}llner},
  \citenamefont {Stobbe},\ and\ \citenamefont
  {Lodahl}}]{mahmoodian2017engineering}%
  \BibitemOpen
  \bibfield  {author} {\bibinfo {author} {\bibfnamefont {S.}~\bibnamefont
  {Mahmoodian}}, \bibinfo {author} {\bibfnamefont {K.}~\bibnamefont
  {Prindal-Nielsen}}, \bibinfo {author} {\bibfnamefont {I.}~\bibnamefont
  {S{\"o}llner}}, \bibinfo {author} {\bibfnamefont {S.}~\bibnamefont
  {Stobbe}},\ and\ \bibinfo {author} {\bibfnamefont {P.}~\bibnamefont
  {Lodahl}},\ }\href {https://doi.org/10.1364/OME.7.000043} {\bibfield
  {journal} {\bibinfo  {journal} {Opt. Mater. Express}\ }\textbf {\bibinfo
  {volume} {7}},\ \bibinfo {pages} {43} (\bibinfo {year} {2017})}\BibitemShut
  {NoStop}%
\bibitem [{\citenamefont {Pedersen}\ \emph {et~al.}(2021)\citenamefont
  {Pedersen}, \citenamefont {González-Ruiz}, \citenamefont {Hauff},
  \citenamefont {Wang}, \citenamefont {Wieck}, \citenamefont {Ludwig},
  \citenamefont {Schott}, \citenamefont {Midolo}, \citenamefont {Sørensen},
  \citenamefont {Uppu},\ and\ \citenamefont {Lodahl}}]{pedersen_-demand_2021}%
  \BibitemOpen
  \bibfield  {author} {\bibinfo {author} {\bibfnamefont {F.~T.}\ \bibnamefont
  {Pedersen}}, \bibinfo {author} {\bibfnamefont {E.~M.}\ \bibnamefont
  {González-Ruiz}}, \bibinfo {author} {\bibfnamefont {N.}~\bibnamefont
  {Hauff}}, \bibinfo {author} {\bibfnamefont {Y.}~\bibnamefont {Wang}},
  \bibinfo {author} {\bibfnamefont {A.~D.}\ \bibnamefont {Wieck}}, \bibinfo
  {author} {\bibfnamefont {A.}~\bibnamefont {Ludwig}}, \bibinfo {author}
  {\bibfnamefont {R.}~\bibnamefont {Schott}}, \bibinfo {author} {\bibfnamefont
  {L.}~\bibnamefont {Midolo}}, \bibinfo {author} {\bibfnamefont {A.~S.}\
  \bibnamefont {Sørensen}}, \bibinfo {author} {\bibfnamefont {R.}~\bibnamefont
  {Uppu}},\ and\ \bibinfo {author} {\bibfnamefont {P.}~\bibnamefont {Lodahl}},\
  }\href {http://arxiv.org/abs/2109.03519} {\bibfield  {journal} {\bibinfo
  {journal} {arXiv:2109.03519 [physics, physics:quant-ph]}\ } (\bibinfo {year}
  {2021})}\BibitemShut {NoStop}%
\bibitem [{\citenamefont {Orazbayev}\ \emph {et~al.}(2018)\citenamefont
  {Orazbayev}, \citenamefont {Kaina},\ and\ \citenamefont
  {Fleury}}]{orazbayev_chiral_2018}%
  \BibitemOpen
  \bibfield  {author} {\bibinfo {author} {\bibfnamefont {B.}~\bibnamefont
  {Orazbayev}}, \bibinfo {author} {\bibfnamefont {N.}~\bibnamefont {Kaina}},\
  and\ \bibinfo {author} {\bibfnamefont {R.}~\bibnamefont {Fleury}},\ }\href
  {https://doi.org/10.1103/PhysRevApplied.10.054069} {\bibfield  {journal}
  {\bibinfo  {journal} {Physical Review Applied}\ }\textbf {\bibinfo {volume}
  {10}},\ \bibinfo {pages} {054069} (\bibinfo {year} {2018})}\BibitemShut
  {NoStop}%
\bibitem [{\citenamefont {Orazbayev}\ and\ \citenamefont
  {Fleury}(2019)}]{orazbayev_quantitative_2019}%
  \BibitemOpen
  \bibfield  {author} {\bibinfo {author} {\bibfnamefont {B.}~\bibnamefont
  {Orazbayev}}\ and\ \bibinfo {author} {\bibfnamefont {R.}~\bibnamefont
  {Fleury}},\ }\href {https://doi.org/10.1515/nanoph-2019-0137} {\bibfield
  {journal} {\bibinfo  {journal} {Nanophotonics}\ }\textbf {\bibinfo {volume}
  {8}},\ \bibinfo {pages} {1433} (\bibinfo {year} {2019})}\BibitemShut
  {NoStop}%
\bibitem [{\citenamefont {Hauff}\ \emph {et~al.}(2021)\citenamefont {Hauff},
  \citenamefont {Hughes}, \citenamefont {Jeannic}, \citenamefont {Lodahl},\
  and\ \citenamefont {Rotenberg}}]{hauff_chiral_2021}%
  \BibitemOpen
  \bibfield  {author} {\bibinfo {author} {\bibfnamefont {N.}~\bibnamefont
  {Hauff}}, \bibinfo {author} {\bibfnamefont {S.}~\bibnamefont {Hughes}},
  \bibinfo {author} {\bibfnamefont {H.~L.}\ \bibnamefont {Jeannic}}, \bibinfo
  {author} {\bibfnamefont {P.}~\bibnamefont {Lodahl}},\ and\ \bibinfo {author}
  {\bibfnamefont {N.}~\bibnamefont {Rotenberg}},\ }\href
  {https://arxiv.org/abs/2111.02828v1} {\bibfield  {journal} {\bibinfo
  {journal} {arXiv:2111.02828 [physics.optics]}\ } (\bibinfo {year}
  {2021})}\BibitemShut {NoStop}%
\bibitem [{\citenamefont {Sotto}\ \emph {et~al.}(2018)\citenamefont {Sotto},
  \citenamefont {Debnath}, \citenamefont {Khokhar}, \citenamefont {Tomita},
  \citenamefont {Thomson},\ and\ \citenamefont {Saito}}]{sotto2018anomalous}%
  \BibitemOpen
  \bibfield  {author} {\bibinfo {author} {\bibfnamefont {M.}~\bibnamefont
  {Sotto}}, \bibinfo {author} {\bibfnamefont {K.}~\bibnamefont {Debnath}},
  \bibinfo {author} {\bibfnamefont {A.~Z.}\ \bibnamefont {Khokhar}}, \bibinfo
  {author} {\bibfnamefont {I.}~\bibnamefont {Tomita}}, \bibinfo {author}
  {\bibfnamefont {D.}~\bibnamefont {Thomson}},\ and\ \bibinfo {author}
  {\bibfnamefont {S.}~\bibnamefont {Saito}},\ }\href
  {https://doi.org/10.1364/JOSAB.35.002356} {\bibfield  {journal} {\bibinfo
  {journal} {J. Opt. Soc. Am. B}\ }\textbf {\bibinfo {volume} {35}},\ \bibinfo
  {pages} {2356} (\bibinfo {year} {2018})}\BibitemShut {NoStop}%
\bibitem [{\citenamefont {Yoshimi}\ \emph {et~al.}(2021)\citenamefont
  {Yoshimi}, \citenamefont {Yamaguchi}, \citenamefont {Katsumi}, \citenamefont
  {Ota}, \citenamefont {Arakawa},\ and\ \citenamefont
  {Iwamoto}}]{yoshimi_experimental_2021}%
  \BibitemOpen
  \bibfield  {author} {\bibinfo {author} {\bibfnamefont {H.}~\bibnamefont
  {Yoshimi}}, \bibinfo {author} {\bibfnamefont {T.}~\bibnamefont {Yamaguchi}},
  \bibinfo {author} {\bibfnamefont {R.}~\bibnamefont {Katsumi}}, \bibinfo
  {author} {\bibfnamefont {Y.}~\bibnamefont {Ota}}, \bibinfo {author}
  {\bibfnamefont {Y.}~\bibnamefont {Arakawa}},\ and\ \bibinfo {author}
  {\bibfnamefont {S.}~\bibnamefont {Iwamoto}},\ }\href
  {http://arxiv.org/abs/2102.09252} {\bibfield  {journal} {\bibinfo  {journal}
  {Opt. Express}\ }\textbf {\bibinfo {volume} {29}},\ \bibinfo {pages} {13441}
  (\bibinfo {year} {2021})}\BibitemShut {NoStop}%
\bibitem [{\citenamefont {Kuang}\ and\ \citenamefont
  {O’Brien}(2004)}]{kuang2004reducing}%
  \BibitemOpen
  \bibfield  {author} {\bibinfo {author} {\bibfnamefont {W.}~\bibnamefont
  {Kuang}}\ and\ \bibinfo {author} {\bibfnamefont {J.~D.}\ \bibnamefont
  {O’Brien}},\ }\href {https://doi.org/10.1364/OL.29.000860} {\bibfield
  {journal} {\bibinfo  {journal} {Opt. letters}\ }\textbf {\bibinfo {volume}
  {29}},\ \bibinfo {pages} {860} (\bibinfo {year} {2004})}\BibitemShut
  {NoStop}%
\bibitem [{\citenamefont {Sotto}\ \emph {et~al.}(2019)\citenamefont {Sotto},
  \citenamefont {Debnath}, \citenamefont {Tomita},\ and\ \citenamefont
  {Saito}}]{sotto2019spin}%
  \BibitemOpen
  \bibfield  {author} {\bibinfo {author} {\bibfnamefont {M.}~\bibnamefont
  {Sotto}}, \bibinfo {author} {\bibfnamefont {K.}~\bibnamefont {Debnath}},
  \bibinfo {author} {\bibfnamefont {I.}~\bibnamefont {Tomita}},\ and\ \bibinfo
  {author} {\bibfnamefont {S.}~\bibnamefont {Saito}},\ }\href
  {https://doi.org/10.1103/PhysRevA.99.053845} {\bibfield  {journal} {\bibinfo
  {journal} {Phys. Rev. A}\ }\textbf {\bibinfo {volume} {99}},\ \bibinfo
  {pages} {053845} (\bibinfo {year} {2019})}\BibitemShut {NoStop}%
\bibitem [{\citenamefont {Mock}(2020)}]{mock2020symmetry}%
  \BibitemOpen
  \bibfield  {author} {\bibinfo {author} {\bibfnamefont {A.}~\bibnamefont
  {Mock}},\ }\href {https://doi.org/10.1364/JOSAB.37.000168} {\bibfield
  {journal} {\bibinfo  {journal} {J. Opt. Soc. Am. B}\ }\textbf {\bibinfo
  {volume} {37}},\ \bibinfo {pages} {168} (\bibinfo {year} {2020})}\BibitemShut
  {NoStop}%
\bibitem [{\citenamefont {Notomi}\ \emph {et~al.}(2001)\citenamefont {Notomi},
  \citenamefont {Yamada}, \citenamefont {Shinya}, \citenamefont {Takahashi},
  \citenamefont {Takahashi},\ and\ \citenamefont
  {Yokohama}}]{notomi2001extremely}%
  \BibitemOpen
  \bibfield  {author} {\bibinfo {author} {\bibfnamefont {M.}~\bibnamefont
  {Notomi}}, \bibinfo {author} {\bibfnamefont {K.}~\bibnamefont {Yamada}},
  \bibinfo {author} {\bibfnamefont {A.}~\bibnamefont {Shinya}}, \bibinfo
  {author} {\bibfnamefont {J.}~\bibnamefont {Takahashi}}, \bibinfo {author}
  {\bibfnamefont {C.}~\bibnamefont {Takahashi}},\ and\ \bibinfo {author}
  {\bibfnamefont {I.}~\bibnamefont {Yokohama}},\ }\href
  {https://doi.org/10.1103/PhysRevLett.87.253902} {\bibfield  {journal}
  {\bibinfo  {journal} {Phys. Rev. Lett.}\ }\textbf {\bibinfo {volume} {87}},\
  \bibinfo {pages} {253902} (\bibinfo {year} {2001})}\BibitemShut {NoStop}%
\bibitem [{\citenamefont {Vlasov}\ \emph {et~al.}(2005)\citenamefont {Vlasov},
  \citenamefont {O'Boyle}, \citenamefont {Hamann},\ and\ \citenamefont
  {McNab}}]{vlasov2005active}%
  \BibitemOpen
  \bibfield  {author} {\bibinfo {author} {\bibfnamefont {Y.~A.}\ \bibnamefont
  {Vlasov}}, \bibinfo {author} {\bibfnamefont {M.}~\bibnamefont {O'Boyle}},
  \bibinfo {author} {\bibfnamefont {H.~F.}\ \bibnamefont {Hamann}},\ and\
  \bibinfo {author} {\bibfnamefont {S.~J.}\ \bibnamefont {McNab}},\ }\href
  {https://doi.org/10.1038/nature04210} {\bibfield  {journal} {\bibinfo
  {journal} {Nature}\ }\textbf {\bibinfo {volume} {438}},\ \bibinfo {pages}
  {65} (\bibinfo {year} {2005})}\BibitemShut {NoStop}%
\bibitem [{\citenamefont {Johnson}\ and\ \citenamefont
  {Joannopoulos}(2001)}]{MPB}%
  \BibitemOpen
  \bibfield  {author} {\bibinfo {author} {\bibfnamefont {S.~G.}\ \bibnamefont
  {Johnson}}\ and\ \bibinfo {author} {\bibfnamefont {J.~D.}\ \bibnamefont
  {Joannopoulos}},\ }\href {https://doi.org/10.1364/OE.8.000173} {\bibfield
  {journal} {\bibinfo  {journal} {Opt. Express}\ }\textbf {\bibinfo {volume}
  {8}},\ \bibinfo {pages} {173} (\bibinfo {year} {2001})}\BibitemShut {NoStop}%
\bibitem [{\citenamefont {Khanikaev}\ and\ \citenamefont
  {Shvets}(2017)}]{khanikaev2017two}%
  \BibitemOpen
  \bibfield  {author} {\bibinfo {author} {\bibfnamefont {A.~B.}\ \bibnamefont
  {Khanikaev}}\ and\ \bibinfo {author} {\bibfnamefont {G.}~\bibnamefont
  {Shvets}},\ }\href {https://doi.org/10.1038/s41566-017-0048-5} {\bibfield
  {journal} {\bibinfo  {journal} {Nat. Photon.}\ }\textbf {\bibinfo {volume}
  {11}},\ \bibinfo {pages} {763} (\bibinfo {year} {2017})}\BibitemShut
  {NoStop}%
\bibitem [{\citenamefont {Faraon}\ \emph {et~al.}(2008)\citenamefont {Faraon},
  \citenamefont {Fushman}, \citenamefont {Englund}, \citenamefont {Stoltz},
  \citenamefont {Petroff},\ and\ \citenamefont
  {Vu{\v{c}}kovi{\'c}}}]{faraon2008dipole}%
  \BibitemOpen
  \bibfield  {author} {\bibinfo {author} {\bibfnamefont {A.}~\bibnamefont
  {Faraon}}, \bibinfo {author} {\bibfnamefont {I.}~\bibnamefont {Fushman}},
  \bibinfo {author} {\bibfnamefont {D.}~\bibnamefont {Englund}}, \bibinfo
  {author} {\bibfnamefont {N.}~\bibnamefont {Stoltz}}, \bibinfo {author}
  {\bibfnamefont {P.}~\bibnamefont {Petroff}},\ and\ \bibinfo {author}
  {\bibfnamefont {J.}~\bibnamefont {Vu{\v{c}}kovi{\'c}}},\ }\href
  {https://doi.org/10.1364/OE.16.012154} {\bibfield  {journal} {\bibinfo
  {journal} {Opt. Express}\ }\textbf {\bibinfo {volume} {16}},\ \bibinfo
  {pages} {12154} (\bibinfo {year} {2008})}\BibitemShut {NoStop}%
\bibitem [{\citenamefont {Lee}\ \emph {et~al.}(2008)\citenamefont {Lee},
  \citenamefont {Grillet}, \citenamefont {Poulton}, \citenamefont {Monat},
  \citenamefont {Smith}, \citenamefont {M{\"a}gi}, \citenamefont {Freeman},
  \citenamefont {Madden}, \citenamefont {Luther-Davies},\ and\ \citenamefont
  {Eggleton}}]{lee2008characterizing}%
  \BibitemOpen
  \bibfield  {author} {\bibinfo {author} {\bibfnamefont {M.~W.}\ \bibnamefont
  {Lee}}, \bibinfo {author} {\bibfnamefont {C.}~\bibnamefont {Grillet}},
  \bibinfo {author} {\bibfnamefont {C.~G.}\ \bibnamefont {Poulton}}, \bibinfo
  {author} {\bibfnamefont {C.}~\bibnamefont {Monat}}, \bibinfo {author}
  {\bibfnamefont {C.~L.}\ \bibnamefont {Smith}}, \bibinfo {author}
  {\bibfnamefont {E.}~\bibnamefont {M{\"a}gi}}, \bibinfo {author}
  {\bibfnamefont {D.}~\bibnamefont {Freeman}}, \bibinfo {author} {\bibfnamefont
  {S.}~\bibnamefont {Madden}}, \bibinfo {author} {\bibfnamefont
  {B.}~\bibnamefont {Luther-Davies}},\ and\ \bibinfo {author} {\bibfnamefont
  {B.~J.}\ \bibnamefont {Eggleton}},\ }\href
  {https://doi.org/10.1364/OE.16.013800} {\bibfield  {journal} {\bibinfo
  {journal} {Opt. Express}\ }\textbf {\bibinfo {volume} {16}},\ \bibinfo
  {pages} {13800} (\bibinfo {year} {2008})}\BibitemShut {NoStop}%
\bibitem [{\citenamefont {Arregui}\ \emph {et~al.}(2018)\citenamefont
  {Arregui}, \citenamefont {Navarro-Urrios}, \citenamefont {Kehagias},
  \citenamefont {Torres},\ and\ \citenamefont
  {García}}]{arregui_all-optical_2018}%
  \BibitemOpen
  \bibfield  {author} {\bibinfo {author} {\bibfnamefont {G.}~\bibnamefont
  {Arregui}}, \bibinfo {author} {\bibfnamefont {D.}~\bibnamefont
  {Navarro-Urrios}}, \bibinfo {author} {\bibfnamefont {N.}~\bibnamefont
  {Kehagias}}, \bibinfo {author} {\bibfnamefont {C.~M.~S.}\ \bibnamefont
  {Torres}},\ and\ \bibinfo {author} {\bibfnamefont {P.~D.}\ \bibnamefont
  {García}},\ }\href {https://doi.org/10.1103/PhysRevB.98.180202} {\bibfield
  {journal} {\bibinfo  {journal} {Phys. Rev. B}\ }\textbf {\bibinfo {volume}
  {98}},\ \bibinfo {pages} {180202} (\bibinfo {year} {2018})}\BibitemShut
  {NoStop}%
\bibitem [{\citenamefont {Sarkissian}\ and\ \citenamefont
  {O'Brien}(2014)}]{sarkissian2014group}%
  \BibitemOpen
  \bibfield  {author} {\bibinfo {author} {\bibfnamefont {R.}~\bibnamefont
  {Sarkissian}}\ and\ \bibinfo {author} {\bibfnamefont {J.}~\bibnamefont
  {O'Brien}},\ }\href {https://doi.org/10.1063/1.4896519} {\bibfield  {journal}
  {\bibinfo  {journal} {Appl. Phys. Lett.}\ }\textbf {\bibinfo {volume}
  {105}},\ \bibinfo {pages} {121102} (\bibinfo {year} {2014})}\BibitemShut
  {NoStop}%
\bibitem [{\citenamefont {Arregui}\ \emph {et~al.}(2021)\citenamefont
  {Arregui}, \citenamefont {Gomis-Bresco}, \citenamefont {Sotomayor-Torres},\
  and\ \citenamefont {Garcia}}]{arregui_quantifying_2021}%
  \BibitemOpen
  \bibfield  {author} {\bibinfo {author} {\bibfnamefont {G.}~\bibnamefont
  {Arregui}}, \bibinfo {author} {\bibfnamefont {J.}~\bibnamefont
  {Gomis-Bresco}}, \bibinfo {author} {\bibfnamefont {C.~M.}\ \bibnamefont
  {Sotomayor-Torres}},\ and\ \bibinfo {author} {\bibfnamefont {P.~D.}\
  \bibnamefont {Garcia}},\ }\href
  {https://doi.org/10.1103/PhysRevLett.126.027403} {\bibfield  {journal}
  {\bibinfo  {journal} {Physical Review Letters}\ }\textbf {\bibinfo {volume}
  {126}},\ \bibinfo {pages} {027403} (\bibinfo {year} {2021})}\BibitemShut
  {NoStop}%
\bibitem [{\citenamefont {Garc{\'\i}a}\ \emph {et~al.}(2010)\citenamefont
  {Garc{\'\i}a}, \citenamefont {Smolka}, \citenamefont {Stobbe},\ and\
  \citenamefont {Lodahl}}]{garcia2010density}%
  \BibitemOpen
  \bibfield  {author} {\bibinfo {author} {\bibfnamefont {P.}~\bibnamefont
  {Garc{\'\i}a}}, \bibinfo {author} {\bibfnamefont {S.}~\bibnamefont {Smolka}},
  \bibinfo {author} {\bibfnamefont {S.}~\bibnamefont {Stobbe}},\ and\ \bibinfo
  {author} {\bibfnamefont {P.}~\bibnamefont {Lodahl}},\ }\href@noop {}
  {\bibfield  {journal} {\bibinfo  {journal} {Phys. Rev. B}\ }\textbf {\bibinfo
  {volume} {82}},\ \bibinfo {pages} {165103} (\bibinfo {year}
  {2010})}\BibitemShut {NoStop}%
\bibitem [{\citenamefont {Mondal}\ \emph {et~al.}(2019)\citenamefont {Mondal},
  \citenamefont {Kumar}, \citenamefont {Kamp},\ and\ \citenamefont
  {Mujumdar}}]{mondal_optical_2019}%
  \BibitemOpen
  \bibfield  {author} {\bibinfo {author} {\bibfnamefont {S.}~\bibnamefont
  {Mondal}}, \bibinfo {author} {\bibfnamefont {R.}~\bibnamefont {Kumar}},
  \bibinfo {author} {\bibfnamefont {M.}~\bibnamefont {Kamp}},\ and\ \bibinfo
  {author} {\bibfnamefont {S.}~\bibnamefont {Mujumdar}},\ }\href
  {https://doi.org/10.1103/PhysRevB.100.060201} {\bibfield  {journal} {\bibinfo
   {journal} {Phys. Rev. B}\ }\textbf {\bibinfo {volume} {100}},\ \bibinfo
  {pages} {060201} (\bibinfo {year} {2019})}\BibitemShut {NoStop}%
\bibitem [{\citenamefont {Edwards}\ and\ \citenamefont
  {Thouless}(1972)}]{edwards1972numerical}%
  \BibitemOpen
  \bibfield  {author} {\bibinfo {author} {\bibfnamefont {J.}~\bibnamefont
  {Edwards}}\ and\ \bibinfo {author} {\bibfnamefont {D.}~\bibnamefont
  {Thouless}},\ }\href@noop {} {\bibfield  {journal} {\bibinfo  {journal} {J.
  Phys. C: Solid State Phys.}\ }\textbf {\bibinfo {volume} {5}},\ \bibinfo
  {pages} {807} (\bibinfo {year} {1972})}\BibitemShut {NoStop}%
\bibitem [{\citenamefont {Mondal}\ and\ \citenamefont
  {Mujumdar}(2020)}]{mondal_relation_2020}%
  \BibitemOpen
  \bibfield  {author} {\bibinfo {author} {\bibfnamefont {S.}~\bibnamefont
  {Mondal}}\ and\ \bibinfo {author} {\bibfnamefont {S.}~\bibnamefont
  {Mujumdar}},\ }\href {https://doi.org/10.1364/OL.383748} {\bibfield
  {journal} {\bibinfo  {journal} {Optics Letters}\ }\textbf {\bibinfo {volume}
  {45}},\ \bibinfo {pages} {997} (\bibinfo {year} {2020})}\BibitemShut
  {NoStop}%
\end{thebibliography}%


%
\end{document}